\documentclass[aps,prb,showpacs,twocolumn,amsmath,amssymb,revsymb,a4paper]{revtex4-1}
\usepackage{graphicx}% Include figure files
\usepackage{color} % color online
\usepackage{dcolumn}% Align table columns on decimal point
\usepackage{bm}% bold math
\usepackage{setspace}
\usepackage{sidecap}
\usepackage{natbib}
\usepackage{hyperref}%makes link in pdf file 

\providecommand{\e}{\varepsilon}
\providecommand{\h}{\hbar}
\providecommand{\p}{\partial}
\newcommand{\kb}{k_{\textsc{b}}^{}}
\newcommand{\kk}{\mathbf{k}}
\newcommand{\op}{\uparrow}
\newcommand{\ned}{\downarrow}
\begin{document}

\title{Helical edge states coupled to a spin bath: Current-induced magnetization}

\author{Anders Mathias Lunde and Gloria Platero}
\affiliation{Instituto de Ciencia de Materiales de Madrid, CSIC, Cantoblanco, 28049 Madrid, Spain}
\date{\today}

\begin{abstract}
We study current carrying helical edge states in a two-dimensional topological insulator coupled to an environment of localized spins, i.e.~a spin bath. The localized spins mediate elastic spin-flip scattering between the helical edge states, and we show how this induces a spin-bath magnetization for a finite current through the edge states. The magnetization appears near the boundaries of the topological insulator, while the bulk remains unmagnetized, and it reaches its maximal value in the high bias regime. Furthermore, the helical edge states remain ballistic in steady state, if no additional spin-flip mechanisms for the localized spins are present. However, we demonstrate that if such mechanisms are allowed, then these will induce a finite current decrease from the ballistic value.
\end{abstract}


\maketitle

\section{Introduction}

Gapless helical edge states exist at the boundary of a two-dimensional (2D) topological insulator (TI)\cite{Qi-Zhang-review-RMP-2010,Hasan-review-RMP-2010,Kane-Mele-first-paper-PRL-2005}. A pair of helical edge states are counter propagating, and different edge states with opposite wave numbers, $k$ and $-k$, constitute a Kramers pair. The prime example of a 2D TI has been recently realized experimentally by the group of Molenkamp\cite{Konig-Science-2007,Roth-Science-2009,Brune-Molenkamp-nature-phys-2012,Konig-JPSJ-review-2008} in HgTe quantum wells. Evidence of edge states in both two-terminal\cite{Konig-Science-2007} and non-local\cite{Roth-Science-2009} transport measurements were found and, furthermore, the relation between the spin of the helical edge state (HES) and propagation direction was experimentally established\cite{Brune-Molenkamp-nature-phys-2012}. The existence of the TI state in a HgTe quantum well beyond a certain critical well thickness was predicted by Bernevig, Hughes and Zhang (BHZ)\cite{Bernevig-Zhang-Science-2006} by constructing a minimal model -- similar to a massive Dirac model -- describing the basic physics.

\begin{figure}
\includegraphics[width=0.42\textwidth,angle=0]{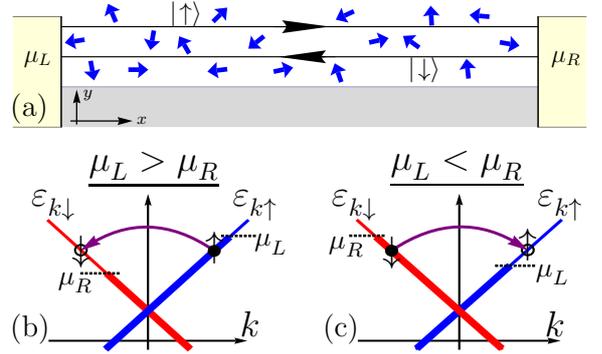}
\caption{(color online) (a) A pair of helical edge states coupled to a spin bath (blue arrows) between two leads with chemical potentials $\mu_{L}$ and $\mu_{R}$, respectively. The pair of helical edge states exists at the interface between a 2D topological insulator (white region) and an ordinary insulator (grey area). (b) and (c) show the dominant spin-flip scattering processes between the edge states for $\mu_{L}>\mu_{R}$ and $\mu_{R}>\mu_{L}$, respectively, for an \emph{unmagnetized} bath. Here the right- (left-) moving edge state is occupied up to $\mu_L$ ($\mu_R$), as indicated by the thick blue (red) lines. Since the inter edge state scattering is mediated by a flip of a localized spin in the bath, it will lead to an excess number of spin up in the bath for $\mu_{L}>\mu_{R}$ in (b) (and vice versa for $\mu_{R}>\mu_{L}$ in (c)), and in terms give a finite magnetization.}
\label{fig:HES-coupled-to-spin-bath}
\end{figure}

The fact that the HESs come in Kramers pairs means that elastic scattering from wave-vector $k$ in one HES to $-k$ in the other HES within a pair cannot be induced by time-reversal invariant potentials. Therefore scattering by e.g. impurities between a pair of HESs is strongly suppressed, which results in ballistic transport and quantized conductance of $e^2/h$ per pair of HESs as observed\cite{Konig-Science-2007,Roth-Science-2009}. Inelastic scattering mechanisms\cite{Budich-Dolcini-Recher-Trauzettel-PRL-2012,Schmidt-Rachel-Oppen-Glazman-PRL-2012,Lezmy-Oreg-Berkooz-PRB-2012} and scattering mechanisms breaking time-reversal invariance can, however, induce scattering between the HESs. For instance, the effect of a single magnetic impurity on the conductance through a pair of HESs have been considered\cite{Maciejko-PRL-2009,Tanaka-Furusaki-Matveev-PRL-2011}. In this case, Tanaka \emph{et al.}\cite{Tanaka-Furusaki-Matveev-PRL-2011} showed that even though a single magnetic impurity breaks time-reversal invariance, it does not have any effect on the dc conductance.
Also the RKKY interaction mediated by HESs has been studied\cite{Gao-et-al-PRB-2009}.

\subsection{Qualitative considerations on the current induced magnetization and the current change}

In this paper, we consider a 2D TI coupled to an environment of localized spins, i.e.~a spin bath. We focus on a single pair of HESs as seen in Fig.~\ref{fig:HES-coupled-to-spin-bath}(a), even though a real device has a pair of HESs at each boundary. This can be done without loss of generality as long as the boundaries are not close together (as for instance in a point contact geometry\cite{Krueckl-Richter-PRL-2011}). Furthermore, for simplicity we consider the case of a spin--1/2 bath, which is not essential for the physics discussed in this paper.

The spin bath breaks time-reversal symmetry (from the point of view of the carriers in the HESs) and therefore enables the possibility of elastic scattering between the HESs. We argue that current through the HESs will induce a magnetization in the spin bath near the boundary region of the 2D TI. The magnetization comes about due to angular momentum conserving scattering between the HESs. The occupation of the HES is determined by the contact, where it originates. This means that the right moving HES is occupied up to the chemical potential $\mu_L$ of the left contact and vice versa (see Fig.~\ref{fig:HES-coupled-to-spin-bath}). Therefore, a finite bias voltage, say $\mu_L>\mu_R$, opens an energy window favoring scattering from the right-moving spin up HES to the left-moving spin down HES, see Fig.~\ref{fig:HES-coupled-to-spin-bath}(b). This scattering is mediated by a spin flip in the bath in the opposite direction, $\ned$ to $\op$, and therefore this rate of inter-HES scattering depends on the number of spin down in the bath. The scattering between the HES will therefore dynamically change the number of spin down and up in the spin bath until a steady state is reached. In other words, the magnetization in the spin bath builds up to compensate the increased phase space for the inter-HES scattering due to the bias. In particular, the spin bath can magnetize completely in the high bias regime, where the phase space for one of the two inter-HES scattering processes is suppressed completely. Since the HESs only exist on the boundary, the spins localized in the bulk of the 2D TI are not affected by the current through the HESs. 

The spin-flip scattering between the HESs also change the propagation direction of the carrier, i.e.~it is a backscattering process. However, if no additional spin-flip mechanisms are present for the localized spins, then the HESs remain ballistic in steady state, because once a spin in the bath has mediated a transition between the edge states by flipping from -say- down to up, then it cannot mediate another transition. Nevertheless, if additional spin-flip mechanisms are feasible to randomize the direction of the localized spins, then this will induce a finite steady state current change, since backscattering between the edge states now will try to compensate the randomizing of the spins. 

One possible experimental realization of the spin bath is magnetic impurities, e.g.~Mn ions in a HgTe quantum well\cite{Novik-Molenkamp-PRB-2005}.  In this case, low concentration of magnetic impurities is required not to hinder the existence of HESs. However, the physics and phenomenon discussed here is of a rather generic nature for any 2D TI coupled to an environment of localized spins. 

\section{The spin-bath dynamics and the current}

Next, we describe in detail how the spin bath and the transport through the HESs are connected. Without the spin bath, the electric current through a single pair of HESs is ballistic such that 
\begin{align}
I^{(0)}=\frac{(-e)}{h}(\mu_L-\mu_R),
\end{align} 
where $\mu_L$ ($\mu_R$) is the chemical potential of the left (right) lead and $e>0$ is the elementary charge. 

The spin-bath mediated scattering between the HESs is a backscattering of a single particle, since the two HESs are counterpropagating. The electric current change $\delta I$ due to the spin bath is given by the rate of change in the number of left (or right) movers. This is, in terms, given by the rates $\Gamma_{\sigma\leftarrow\sigma'}$ for transferring a particle from the HES $\sigma'$ to the HES $\sigma$, i.e.    
\begin{align}
\delta I=(-e)(\Gamma_{\op\leftarrow\ned}-\Gamma_{\ned\leftarrow\op}),
\label{eq:general-current-change} 
\end{align}
and the total current is $I=I^{(0)}+\delta I$. The scattering rates $\Gamma_{\sigma\leftarrow\sigma'}$ depend on the magnetization of the spin bath. For instance, the more localized spins with spin up, the larger the rate $\Gamma_{\op\leftarrow\ned}$ and vice versa. (Detailed expressions are given below, see e.g.~Eq.(\ref{eq:HES-rates-general}).) 

Now we turn to the dynamics of the spin bath. Every time one particle is scattered between the HESs, a single localized spin is flipped. This means that the number of each spin species in the bath covered by the HES, $N_{\sigma}$, will change dynamically with the same rates $\Gamma_{\sigma\leftarrow\sigma'}$ as the inter-HES scattering. Therefore, the time evolution of the number of localized spin species $\sigma$, $N_{\sigma}$, due to the scattering between the HESs can be described by the rate equations $\p_tN_{\op}=\Gamma_{\ned\leftarrow\op}-\Gamma_{\op\leftarrow\ned}$, and $\p_tN_{\ned}=\Gamma_{\op\leftarrow\ned}-\Gamma_{\ned\leftarrow\op}$, where the total number of localized spins covered by the HESs, $N_s=N_{\op}+N_{\ned}$, is fixed. (Note that the rate indices refer to the HES spin flip, which is opposite to the spin flip in the bath.) For convenience, we normalize the magnetization such that the maximal (minimal) magnetization, where all the localized spins are in the up (down) state, is $1$ ($-1$). Therefore, we write the magnetization as $\mathcal{M}\equiv(N_\op-N_\ned)/N_s$, which is also often called magnetic polarization. Therefore, subtracting the two rate equations for $N_\op$ and $N_\ned$, the rate equation for the magnetization becomes $\p_t\mathcal{M}=2\big(\Gamma_{\ned\leftarrow\op}-\Gamma_{\op\leftarrow\ned}\big)/N_s$. However, this is considering \emph{only} the spin flip in the bath stemming from the scattering between the HESs. Other -- presumably much weaker -- mechanisms might also flip the localized spins such as dipole-dipole interactions within the spin bath and spin-phonon coupling e.g.~relevant for Mn ions\cite{Cao-Besombes-FernandezRossier-PRB-2011,Chudnovsky-Garanin-Schilling-PRB-2005}. Such mechanisms will try to equilibrate the number of spin up and down in the bath, and thus drive the magnetization towards zero. We include this in the time evolution of the magnetization by a phenomenological term $-\Gamma_r\mathcal{M}/N_s$ similar to the relaxation-time approximation\cite{Ashcroft-Mermin-BOOK}, i.e.  
\begin{align}
\p_t\mathcal{M}=\frac{2}{N_s}\big(\Gamma_{\ned\leftarrow\op}-\Gamma_{\op\leftarrow\ned}\big)
-\frac{1}{N_s}\Gamma_r\mathcal{M}.
\label{eq:rate-eq-pol-with-Gamma-r}
\end{align} 
The phenomenological term is divided by $N_s$ such that $\Gamma_r$ is a spin-flip rate per localized spin and thereby comparable to $\Gamma_{\sigma\leftarrow\sigma'}$.

Therefore, it is now evident that for $\Gamma_r=0$, these simple  rate equations lead to $\delta I(t)=\frac{e}{2}N_s\p_t\mathcal{M}$, such that in steady state, $\p_t\mathcal{M}=0$, there is no current change, 
\begin{align}
\delta I=0 
\qquad (\textrm{for}\ \Gamma_r=0),
\end{align}
and the HESs remain ballistic. Physically, the magnetization builds up to compensate the difference in scattering rates between the two HESs (i.e.~$\Gamma_{\ned\leftarrow\op}=\Gamma_{\op\leftarrow\ned}$ is required in steady state for $\Gamma_r=0$).

Taking the additional weak spin-flip mechanisms in the bath into account, $\Gamma_r\neq 0$, the current change is no longer zero in the steady state, but found to be
\begin{align}
\delta I=\frac{e}{2}\Gamma_r\mathcal{M}
\neq0
\label{eq:current-change-general}
\end{align}
by inserting the current change (\ref{eq:general-current-change}) into $\p_t\mathcal{M}=0$. Physically, the current change is a result of the competition between the additional spin flip mechanisms within the bath and the spin flips due to the inter-HES scattering. Below, the scattering rates are found such that the magnetization and current change can be studied in greater detail.

\section{The helical edge states and their coupling to a spin bath}

We model the HESs by the eigenstates 
\begin{subequations}
\label{eq:HES-eigenstates}
\begin{align}
\varphi_{k\op}(x,y)&=\frac{1}{\sqrt{L}}e^{ikx}f_k(y)|\op\rangle,\\
\varphi_{k\ned}(x,y)&=\frac{1}{\sqrt{L}}e^{ikx}f_{-k}(y)|\ned\rangle,
\end{align}
\end{subequations} 
where $f_{k}(y)$ is the (real) transverse wavefunction of width $W_y$ localized at the boundary of the 2D TI and $L$ is the length of the HES, see Fig.\ref{fig:HES-coupled-to-spin-bath}(a). The energies are $\e_{k\sigma}=\e_0+\mathfrak{s}\h v_0 k$, where $\mathfrak{s}=+1 (-1)$ is for spin $\sigma=\op(\ned)$, $v_0>0$ is the velocity and $\e_0$ a constant energy shift. Therefore, spin $\op$ is right moving ($v_{k\op}=\p_k\e_{k\op}/\h=v_0>0$) and spin $\ned$ is left moving. The states $\varphi_{k\op}$ and $\varphi_{-k\ned}$ form a Kramers pair, since $\Theta\varphi_{k\op}=+\varphi_{-k\ned}$ and $\Theta\varphi_{-k\ned}=-\varphi_{k\op}$, where $\Theta=-i\sigma_yK$ is the time-reversal operator consisting of a Pauli matrix $\sigma_y$ and a complex conjugation operator $K$. It is possible to find a specific form of the HESs within the BHZ model, see Refs.~\onlinecite{Zhou-edge-states-PRL-2008}, \onlinecite{Bihlmayer-Edge-states-in-Bi-films-PRB-2011} and Appendix \ref{HES-in-BHZ-model-appendix}. Within this model, each of the two HESs in the Kramers pair consists of a mixture of two orbital states both with either positive or negative total angular momentum projection. Thus, it is possible to model the HESs as spin--$1/2$, which is sufficient for the present purpose. 

The coupling of the HESs to the spin bath is modeled as being point-like both along and transverse to the HES\cite{Culcer-Das-Sarma-PRB-2011,Gao-et-al-PRB-2009,Zhu-et-al-Zhang-PRL-2011,Jiang-Wu-PRB-2011}, i.e.~$V(x,y)=\frac{Ja}{\h^2}\sum_j\delta(x-X_j)\delta(y-Y_j)\mathbf{s}\cdot\mathbf{S}^j$, where $J$ is the coupling (energy) constant, $a$ is the area covered by a single spin $\mathbf{S}^{j}=(S^j_{x},S^j_{y},S^j_{z})$ at the fixed position $(X_j,Y_j)$ and $\mathbf{s}=(s_{x},s_{y},s_{z})$ is the HES spin. Using the eigenstates in Eq.(\ref{eq:HES-eigenstates}), the interaction with the spin bath becomes
\begin{align}
V\!=&\sum_{k,k'}\sum_j e^{i(k'-k)X_j}\!
\Big[
\!J^j_{k\op,k'\op}S^j_z c^{\dag}_{k\op}c^{}_{k'\op}
\!-\!J^j_{k\ned,k'\ned}S^j_z c^{\dag}_{k\ned}c^{}_{k'\ned}
\nonumber\\
&+J^j_{k\op,k'\ned}S^j_{-} c^{\dag}_{k\op}c^{}_{k'\ned}
+J^j_{k\ned,k'\op}S^j_{+} c^{\dag}_{k\ned}c^{}_{k'\op}
\Big], 
\label{eq:spin-bath-coupling-general}
\end{align}
where $S^j_{\pm}=S^j_{x}\pm iS^j_{y}$ are the raising and lowering operators of the $j^{\textrm{th}}$ localized spin and $c^{\dag}_{k\sigma}$ ($c^{}_{k\sigma}$) is the creation (annihilation) operator of the HES $\varphi_{k\sigma}$. The position and wave-vector-dependent coupling is $J^j_{k\sigma,k'\sigma'}=\frac{J}{2\h}\frac{a}{L}f_{\mathfrak{s}k}(Y_j)f_{\mathfrak{s}'k'}(Y_j)$, where $\mathfrak{s}=+1 (-1)$ for $\sigma=\op(\ned)$. The interaction consist of terms representing two kinds of scattering from $k'$ to $k$ of the $j^{\textrm{th}}$ localized spin: (i) Scattering \emph{within} a HES leaving the spin bath unchanged (the two first terms) and (ii) scattering \emph{between} the HESs by flipping a localized spin (the two last terms). It is the second kind of terms, which induce elastic scattering between the HESs from $k$ to $-k$. 

The factor $e^{i(k'-k)X_j}$ in Eq.(\ref{eq:spin-bath-coupling-general}) stems from the point-like nature of the interaction. If the spin-independent part of the potential was taken to be more extended in space, then the factor $e^{i(k'-k)X_j}$ would be replaced by the Fourier transform of the potential in the $x$ direction at $k'-k$. Likewise a broadening in the $y$ direction would introduce an integral over $y$ in the coupling matrix elements $J^j_{k\sigma,k'\sigma'}$. However, these complications are not of importance for the basic physics of the elastic backscattering between the HESs from $k$ to $-k$ discussed here, but could change their effectiveness. Furthermore, here an isotropic interaction between the HES spins $\mathbf{s}$ and the localized spins $\mathbf{S}^j$ is used for simplicity. However, our results are not affected, if uniaxial anisotropy\cite{Maciejko-PRL-2009,Tanaka-Furusaki-Matveev-PRL-2011,Liu-Zhang-PRL-2009} is allowed [such that the coupling constants $J$ would be different in the $z$ direction and in the $(x,y)$ plane].

In passing,  we note that a small energy gap can open up in the HES spectrum due to the spin bath. A way to realize this, is by averaging over the positions of the localized spins in $V$, such that translational invariance is restored. Treating the sum of all the localized spins as a classical field, the Hamiltonian can be diagonalized and it becomes evident that the gap is basically proportional to the in-plane field. 

\section{Inter edge state scattering rates}

For weak coupling to the spin bath, the elastic inter-HES scattering rates $\Gamma_{\sigma\leftarrow\sigma'}$ can be found by the Fermi golden rule to be
%\begin{subequations}
\begin{align}
\Gamma_{\ned\leftarrow\op}&=2\pi \h
\frac{N_\ned}{N_s}
\sum_{kk' j}
|J^j_{k\ned,k'\op}|^2
n_{k'\op}(1-n_{k\ned})
\delta(\e_{k\ned}-\e_{k'\op}), \nonumber\\
\Gamma_{\op\leftarrow\ned}&=2\pi \h
\frac{N_\op}{N_s}
\sum_{kk' j}
|J^j_{k\op,k'\ned}|^2
n_{k'\ned}(1-n_{k\op})
\delta(\e_{k\op}-\e_{k'\ned}),
\label{eq:HES-rates-general}
\end{align}
%\end{subequations}
where $n_{k\sigma}$ is the electronic distribution function of the HES $\sigma$. A detailed derivation is given in Appendix \ref{appendix:HES-scattering-rate}. The physical intuition behind these rates is clear:~A scattering from, say, the HES $\op$ to $\ned$ requires (i) an occupied state to scatter from, $\propto n_{k'\op}$, and (ii) an empty state to scatter into, $\propto 1-n_{k\ned}$. The scattering process conserves energy, i.e.~$\e_{k\ned}=\e_{k'\op}$. Moreover, the inter-HES spin-flip scattering $\op\rightarrow\ned$ demands the presence of a spin down in the bath to mediate the spin flip, which leads to the factor $N_\ned/N_s$.

We observe that inserting these rates into the current Eq.(\ref{eq:general-current-change}), the result is consistent with the Boltzmann equation approach\cite{Lunde-PRL-2006,Jerome-Micklitz-Matveev-PRL-2009,Lunde-PRB-2007,Levchenko-Micklitz-Rech-Matveev-PRB-2010} for general distributions $n_{k\sigma}$. If the time it takes to move through the HES $L/v_0$ (the traversal time), is much shorter than the spin-flip time (inverse spin-flip rate), then the distributions of the HESs $n_{k\sigma}$ are approximately equal to the lead distributions at which they originate, i.e. 
\begin{align}
n_{k\op}\simeq f^0_L(\e_{k\op}) 
\qquad\textrm{and}\qquad 
n_{k\ned}\simeq f^0_R(\e_{k\ned}), 
\label{eq:lead-distributions}
\end{align}
where the leads are Fermi distributed $f^0_{\alpha}(\e)=\{1+\exp[(\e-\mu_{\alpha})/\kb T]\}^{-1}$ for $\alpha=L,R$ and $T$ is the temperature. Therefore, using $N_\op/N_s=\frac{1}{2}(1+\mathcal{M})$ and $N_\ned/N_s=\frac{1}{2}(1-\mathcal{M})$, we end up with
\begin{align}
\Gamma_{\op\leftarrow\ned}
=(1+\mathcal{M})\Gamma_{\op\leftarrow\ned}^0, 
\quad 
\Gamma_{\ned\leftarrow\op}
=(1-\mathcal{M})\Gamma_{\ned\leftarrow\op}^0,
\label{eq:P-dependence-of-rates} 
\end{align}
where the rates at zero magnetization $\Gamma_{\sigma\leftarrow\sigma'}^0$ are
\begin{subequations}
\label{eq:lowest-order-in-L-inter-HES-rates}
\begin{align}
\Gamma^{0}_{\ned\leftarrow\op}&=
\frac{\h L\nu}{2}
\int\!d k
\sum_j 
|J^j_{-k\ned,k\op}|^2
f^0_L(\e_{k\op}^{})[1-f^0_R(\e^{}_{k\op})],\\ 
\Gamma^{0}_{\op\leftarrow\ned}&=
\frac{\h L\nu}{2}
\int\!d k
\sum_{j}
|J^j_{k\op,-k\ned}|^2
f^0_R(\e_{k\op}^{})[1-f^0_L(\e^{}_{k\op})].
\end{align}
\end{subequations}
Here $\nu=L/(2\pi\h v_0)$ is the density of states and $|J^j_{-k\ned,k\op}|^2=|J^j_{k\op,-k\ned}|^2=\left[Ja/(2\h L)\right]^2[f_{+k}(Y_j)]^4$. Due to the energy window of the Fermi functions $f^0_{\alpha'}[1-f^0_\alpha]$, it is now explicitly clear that for $\mathcal{M}=0$ scattering from the HES $\op$ to $\ned$ dominates for $\mu_L^{}>\mu_R^{}$ and vice versa as illustrated in Figs.~\ref{fig:HES-coupled-to-spin-bath}(b) and \ref{fig:HES-coupled-to-spin-bath}(c).

\section{The magnetization and current change}

\begin{figure} 
\includegraphics[width=0.35\textwidth,angle=0]{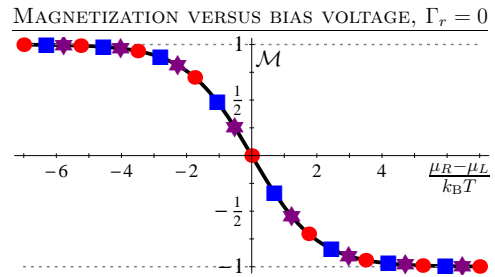}
\caption{(color online) The current-induced magnetization of the localized spins versus bias voltage over temperature, $(\mu_R-\mu_L)/\kb T$, without additional spin-flip mechanisms in the bath, i.e.~$\Gamma_r=0$. The simple approximation (\ref{eq:pol-tanh}) for the magnetization (full black line) is compared to a numerical calculation using the transverse wavefunction from the BHZ model with $N_s=10^3$ randomly chosen spin positions $Y_j$ for $\e_F=(\mu_L+\mu_R)/2=0$ and a temperature of 0.1K (purple stars), 1K (red dots) and 10K (blue squares).}
\label{fig:pol-vs-eV}
\end{figure}

Now the steady state magnetization $\mathcal{M}_{st}$ is easily found by inserting the rates Eq.(\ref{eq:P-dependence-of-rates}) into Eq.(\ref{eq:rate-eq-pol-with-Gamma-r}) and solving $\p_t\mathcal{M}=0$, i.e.
\begin{align}
\mathcal{M}_{st}=
\frac{\Gamma_{\ned\leftarrow\op}^{0}-\Gamma_{\op\leftarrow\ned}^{0}}{\Gamma_{\ned\leftarrow\op}^{0}+\Gamma_{\op\leftarrow\ned}^{0}+\frac{1}{2}\Gamma_r}.
\label{eq:pol-most-general}
\end{align} 
Furthermore, Eq.(\ref{eq:rate-eq-pol-with-Gamma-r}) gives that the magnetization builds up from being initially zero as $\mathcal{M}(t)=\mathcal{M}_{st}(1-e^{-t/\tau^{}_m})$, where $\tau^{}_m=N_s[2(\Gamma_{\ned\leftarrow\op}^{0}+\Gamma_{\op\leftarrow\ned}^{0})+\Gamma_r]^{-1}$ is the characteristic time scale for the magnetization process. Note that while magnetization builds up (i.e.~in the non-stationary regime), $t\lesssim \tau^{}_m$, electrons are backscattered, causing a finite transient current change even for $\Gamma_r=0$ --- in contrast to the stationary regime.  

Neglecting the weak additional spin-flip mechanisms in the bath, $\Gamma_r=0$, the  magnetization is readily obtained from the rates $\Gamma_{\sigma\leftarrow\sigma'}^{0}$ in Eq.(\ref{eq:lowest-order-in-L-inter-HES-rates}) as
\begin{align}
\mathcal{M}_{st}=\frac{\sum_j 
\int d k
|f_{k}(Y_j)|^4
\big[f^0_L(\e_{k\op}^{})-f^0_R(\e^{}_{k\op})\big]}{\sum_j 
\int d k
|f_{k}(Y_j)|^4\ \mathbb{F}(\e^{}_{k\op})},
\label{eq:pol-gammar-zero}
\end{align}
where $\mathbb{F}(\e)\equiv f^0_L(\e)[1-f^0_R(\e)]+f^0_R(\e)[1-f^0_L(\e)]$ was introduced. In the limit of bias voltage $\mu_R-\mu_L$ and temperature $\kb T$ much smaller than the energy variation of the transverse eigenstate, the function $|f_{k}(Y_j)|^4$ can be taken outside the integrals in Eq.(\ref{eq:pol-gammar-zero}), such that it simplifies to
\begin{align}
\mathcal{M}_{st}=\tanh\left(\frac{\mu_L-\mu_R}{2\kb T}\right).
\label{eq:pol-tanh}
\end{align} 
Figure \ref{fig:pol-vs-eV} shows that this is a very good approximation for a HgTe TI. Interestingly, Eq.(\ref{eq:pol-tanh}) resembles the well-known expression of a thermal equilibrium ensemble of spins in an external magnetic field\cite{Kittel-Kroemer-BOOK}, if the bias is exchanged by the Zeeman energy. In contrast, Eq.(\ref{eq:pol-tanh}) describes the \emph{current-induced} magnetization at zero external magnetic field, i.e.~a non-equilibrium steady-state situation. From Eq.(\ref{eq:pol-tanh}) it follows that for bias much larger than temperature, maximal magnetization $\mathcal{M}_{st}\simeq\pm1$ is found. Furthermore, in linear response $|\mu_L-\mu_R|\ll \kb T$, we have $\mathcal{M}_{st}\simeq(\mu_L-\mu_R)/(2\kb T)$.

To test the validity of the approximation (\ref{eq:pol-tanh}), it is compared to a numerical calculation of the magnetization Eq.(\ref{eq:pol-gammar-zero}) in Fig.~\ref{fig:pol-vs-eV}. To this end, we use the transverse state $f_k(y)$ from the BHZ model and randomly chosen spin positions $Y_j$ (see Appendix \ref{HES-in-BHZ-model-appendix} for details on $f_k(y)$). For the parameters for a $70$\AA~wide HgTe quantum well\cite{Qi-Zhang-review-RMP-2010}, the magnetization expression (\ref{eq:pol-tanh}) is found to be an excellent approximation for $0<T<50$K and $N_s$ from ten and up, see Fig.~\ref{fig:pol-vs-eV}. Numerically, the magnetization is also found to be independent of the Fermi level $\e_F\equiv(\mu_L+\mu_R)/2$.   

Next, we turn to the case of including a weak additional spin-flip mechanism, $\Gamma_r\neq0$. For increasing $|\mu_R-\mu_L|/\kb T$ beyond one, Eq.(\ref{eq:lowest-order-in-L-inter-HES-rates}) clearly shows that one of the rates $\Gamma^0_{\sigma\leftarrow\sigma'}$ will increase while the other go to zero. Thus, for $\Gamma_r\neq0$, it is still possible to achieve $\mathcal{M}_{st}\simeq1$ ($-1$)  for $\Gamma^0_{\ned\leftarrow\op}\gg\Gamma_r$ ($\Gamma^0_{\op\leftarrow\ned}\gg\Gamma_r$), which is a stronger requirement than $|\mu_R-\mu_L|\gg \kb T$ as in the $\Gamma_r=0$ case Eq.(\ref{eq:pol-tanh}). Therefore, the maximal possible current change Eq.(\ref{eq:current-change-general}) is $\delta I=\pm e\Gamma_r/2$ for $\mathcal{M}_{st}\rightarrow\pm1$. 

The sample specific information about the positions $(X_j,Y_j)$ of the localized spins is not of importance. Therefore, we introduce the position average of a quantity $A$ as $\bar{A}\equiv\frac{1}{(LW_y)^{N_s}}\int dX_1dY_1\cdots dX_{N_s}dY_{N_s}A$ in analogue with impurity averaging\cite{Flensberg-BOOK}. In other words, it is assumed equally likely to find a localized spin $j$ everywhere in the area covered by the HESs. This enables us to give simpler expressions for magnetization and current change for $\Gamma_r\neq0$. The position averaged inter-HES scattering rate for $\mathcal{M}=0$ is  
\begin{align}
\bar{\Gamma}^{0}_{\ned\leftarrow\op}=\frac{\eta}{\h}(\mu_R-\mu_L)n_B(\mu_R-\mu_L),
\label{eq:position-averaged-rate}
\end{align}
where $n_B(\e)=[e^{\e/\kb T}-1]^{-1}$ is the Bose function and $\eta=(\pi/4)(\nu J)^2 N_s/N^2$ is a dimensionless constant related to the strength of the interaction. Here $N\equiv LW_y/a$ is the number of atoms covered in the plane by the HESs, since both an atom and a localized spin is taken to cover an area of size $a$. To obtain this, we used $\int_0^{\infty} dY_j[f_k(Y_j)]^4\simeq 1/W_y$ (see Appendix \ref{HES-in-BHZ-model-appendix}). For simplicity, the possibility of a weak $k$ dependence of the width $W_y$ is neglected, which is justified within the BHZ model in Appendix \ref{HES-in-BHZ-model-appendix}. The opposite rate $\bar{\Gamma}^{0}_{\op\leftarrow\ned}$ is found by interchanging $\mu_L$ and $\mu_R$ in $\bar{\Gamma}^{0}_{\ned\leftarrow\op}$ Eq.(\ref{eq:position-averaged-rate}). Therefore, using the position averaged rates, the magnetization Eq.(\ref{eq:pol-most-general}) becomes 
\begin{align}
\mathcal{M}_{st}=\frac{\mu_L-\mu_R}{(\mu_L-\mu_R)\coth\left(\frac{\mu_L-\mu_R}{2\kb T}\right)+\frac{\h}{2\eta}\Gamma_r}.
\label{eq:pol-position-averaged}
\end{align}  
Thus, maximal magnetization is accessible for $|\mu_L-\mu_R|\gg \frac{\h}{2\eta}\Gamma_r$ and $|\mu_L-\mu_R|\gg \kb T$. The current change is readily found from Eq.(\ref{eq:current-change-general}) to be  
\begin{align}
\delta I=
-\frac{e^2}{h}V
\frac{\pi\h\Gamma_r}{eV\coth\left(\frac{eV}{2\kb T}\right)+\frac{\h}{2\eta}\Gamma_r},
\end{align} 
where the bias voltage $V\equiv (\mu_R-\mu_L)/e$ was introduced such that $I^{(0)}=\frac{e^2}{h}V$. Therefore, in linear response the correction to the ballistic conductance $G^{(0)}=e^2/h$ is 
\begin{align}
\delta G=
-\frac{e^2}{h}
\frac{\pi\h\Gamma_r}{2\kb T+\frac{\h}{2\eta}\Gamma_r},
\end{align} 
i.e.~both current and conductance are found to decrease compared to the case without a spin bath and furthermore vanish for $\Gamma_r=0$ as expected. 

\section{Discussion and summary}

In this paper, we have discussed the current-induced magnetization at a single boundary of a 2D TI. For a two terminal device with two well-separated boundaries, the spin structure of the HESs is reversed on opposite boundaries (i.e.~if spin-up is right moving on the lower edge of the sample then it is left moving on the upper edge and vice versa). This means that the current-induced magnetization has opposite signs -- but equal magnitudes -- on the two opposite boundaries. The bulk region remains non-magnetic and therefore, the sample as a whole is not magnetized for a two terminal symmetric setup (i.e.~opposite edges have the same length and spin concentration). However, by adjusting the geometry, making the environmental spin concentration inhomogeneous or by using a multi-terminal sample an overall non-zero magnetization can indeed be engineered. A possible application of the current-induced magnetization described in this paper, could be an all electrically accessible memory device.

Two recent works\cite{Rosenberg-Franz-PRB-2012,Lasia-Brey-arxiv-2012} on 3D TIs doped with magnetic impurities discuss how the surface of the TI can magnetize, while the bulk remains non-magnetic (for a certain range of temperatures). In these studies, the magnetic ordering is caused by lowering the temperature leading to a phase transition in an equilibrium setup -- in contrast to the present work describing \emph{current-induced} magnetization. Nevertheless, both effects build on the special nature of the edge and surface states in TIs, where the momentum and spin are locked together. 

Next, we estimate the magnetization time $\tau^{}_m$ for magnetic impurities\cite{Novik-Molenkamp-PRB-2005,Cao-Besombes-FernandezRossier-PRB-2011} to be on the order of tens of $\mu$s for $T=2$K, $\mu_R-\mu_L\sim\kb T$, $Ja\sim1\textrm{eV}$\AA$^2$ and neglecting $\Gamma_r$. Here the velocity $v_0\simeq4\times10^5\textrm{m}/\textrm{s}$  and width $W_y\sim 40$nm correspond to a 70\AA~wide HgTe quantum well within the BHZ model\cite{Qi-Zhang-review-RMP-2010}. Note that $\tau^{}_m$ is independent of the HES length $L$ and concentration of spin $N_s/N$. If the localized spins are the nuclear spins of the 2D TI, then $\tau_{m}^{}$ is much longer since $J$ is much smaller. Therefore, the physics discussed in this paper is more relevant for magnetic impurities embedded in a 2D TI. In both cases, the magnetization time is much longer than the traversal time: $L/v_0\sim10$ps for $L\sim1\mu$m. Furthermore, these estimates indicate that the traversal time is much shorter than the spin-flip time $\sim\tau^{}_m/N_s$ for low concentrations $N_s/N\sim10^{-2}$, such that inserting the lead distributions Eq.(\ref{eq:lead-distributions}) into the rates Eq.(\ref{eq:HES-rates-general}) was indeed a justified approximation. 

In summary, we have shown how a current through the HESs of a 2D TI can induce a magnetization of localized spins embedded in the TI. We have explained this by a simple physical picture of spin-flip scattering between the HESs. Only the region covered by the HESs magnetize. We have demonstrated that if the spin bath is only affected by the inter-HES scattering, then the system remains ballistic. In contrast, if an additional weak spin-flip mechanism is present in the bath, then a finite current decrease is found due to a competition between this mechanism and the spin-flip scattering between the HES.

\section*{Acknowledgments}
We thank Karsten Flensberg, Laurens W. Molenkamp, Sigmund Kohler, Fernando Dominguez and Luis Brey for useful discussions. AML acknowledges the Juan de la Cierva program (MICINN) and we both acknowledge Grant No.~MAT2011-24331 and the ITN Grant 234970(EU).

\appendix 

\section{Derivation of the inter helical edge state scattering rates}\label{appendix:HES-scattering-rate}

In this Appendix, we give a detailed derivation of the elastic scattering rate Eq.(\ref{eq:HES-rates-general}) $\Gamma_{\sigma\leftarrow\sigma'}$ of transferring an electron from the HES with spin $\sigma'$ to the HES with spin $\sigma$. In the limit of weak coupling to the spin bath, it can be found by the Fermi golden rule (see e.g.~Ref.~\onlinecite{Flensberg-BOOK}), 
\begin{align}
\Gamma_{\sigma\leftarrow\sigma'}=
\frac{2\pi}{\h}\sum_{f_{\sigma},i_{\sigma'}}
|\langle f_{\sigma}|V|i_{\sigma'}\rangle|^2
W_{i_{\sigma'}}
\delta(E_{f_{\sigma}}-E_{i_{\sigma'}}).
\label{eq:FGR-general}
\end{align} 
Here the initial state $|i_{\sigma'}\rangle$ (and final state $|f_{\sigma}\rangle$) consist of both an electronic part $|i_{e,\sigma'}\rangle$ and a part for the spin bath $|i_{S}\rangle$, i.e.~$|i_{\sigma'}\rangle=|i_{e,\sigma'}\rangle \otimes |i_{S}\rangle$. The $W_{i_{\sigma'}}$ is the occupation factor of initial states\cite{Flensberg-BOOK}, and $E_{f_{\sigma}}-E_{i_{\sigma'}}$  is the total energy difference between the final and initial state. 

As a specific example, we calculate the rate $\Gamma_{\ned\leftarrow\op}$. To this end, we note that the final state is given in terms of the initial state as $|f_{\ned}\rangle=c^{\dag}_{k'_1\ned}c^{}_{k_1\op}|i_{e,\op}\rangle\otimes\frac{1}{\h}S^{j'}_+|i_{S}\rangle$, i.e. an electron is transferred from $k_1\op$ to $k'_1\ned$ at the cost of a spin flip of the $j'^{\textrm{th}}$ localized spin. Therefore the sum over final states in Eq.(\ref{eq:FGR-general}) becomes a sum over $k_1$, $k'_1$ and $j'$. The occupation factor is written as a product of the electronic and spin-bath part, $W_{i_{\op}}=W_{i_{e,\op}}W_{i_S}$, such that 
\begin{align}
\Gamma_{\ned\leftarrow\op}&=\frac{2\pi}{\h}
\sum_{\substack{k_1k'_1\\kk'}}
\sum_{\substack{jj'\\i_{e,\op}i_{S}}}
|J^j_{k\ned,k'\op}|^2
\frac{1}{\h^2}|\langle i_{S}|S^{j'}_{-}S^j_{+} |i_{S}\rangle|^2
W_{i_S}
\nonumber\\
&\hspace{-0.6cm}\times
|\langle i_{e,\op}| c^{\dag}_{k_1\op}c^{}_{k'_1\ned} c^{\dag}_{k\ned}c^{}_{k'\op}|i_{e,\op}\rangle|^2
W_{i_{e,\op}}
\delta(E_{f_{\ned}}-E_{i_{\op}}),
\end{align}
by inserting the interaction between the localized spins and the HESs Eq.(\ref{eq:spin-bath-coupling-general}) into the Fermi golden rule Eq.(\ref{eq:FGR-general}). The electronic part of the sum gives
\begin{align}
\sum_{i_{e,\op}}
|\langle i_{e,\op}|& c^{\dag}_{k_1\op}c^{}_{k'_1\ned} c^{\dag}_{k\ned}c^{}_{k'\op}|i_{e,\op}\rangle|^2
W_{i_{e,\op}}
\nonumber\\
&=\delta_{k_1,k'}
\delta_{k'_1,k}
n_{k'\op}(1-n_{k\ned}),
\end{align}
where $n_{k\sigma}$ is the electronic distribution function of the HES with spin $\sigma$. The initial state of the spin bath is $|i_{S}\rangle=|m_1,\ldots, m_{N_s}\rangle$, where $m_j=\pm1/2$ is the spin  state of the $j^{\textrm{th}}$ localized spin. Therefore writing the occupation factor $W_{i_S}$ as a product over all the localized spins, $W_{i_S}=W_{m_1}\cdots W_{m_{N_s}}$, we obtain
\begin{align}
\frac{1}{\h^2}
\sum_{i_{S}}
|\langle i_{S}|S^{j'}_{-}S^j_{+} |i_{S}\rangle|^2
W_{i_S}
=
\delta_{j,j'} \h^2 W_{m_j=\ned}, 
\end{align}
where $W_{m_j=\ned}$ is the probability that the $j^{\textrm{th}}$ spin is in the state $m_j=-1/2$. Here we used that the probabilities sum to one, i.e.~$W_{m_i=\ned}+W_{m_i=\op}=1$ for all $i$. We take the probability $W_{m_j=\ned}$ to be $W_{m_j=\ned}=N_\ned/N_s$, where $N_\sigma$ is the number of spins in the bath covered by the HESs with spin $\sigma$  and $N_s=N_\ned+N_\op$. Therefore, we end up with the rate
\begin{align}
\Gamma_{\ned\leftarrow\op}&=2\pi \h
\frac{N_\ned}{N_s}
\sum_{kk' j}
|J^j_{k\ned,k'\op}|^2
n_{k'\op}(1-n_{k\ned})
\delta(\e_{k\ned}-\e_{k'\op}). 
\nonumber
\end{align}
The opposite scattering rate $\Gamma_{\op\leftarrow\ned}$ is found similarly and these are the results used in the main text in Eq.(\ref{eq:HES-rates-general}). 

\section{On the transverse eigenstate in the BHZ model}\label{HES-in-BHZ-model-appendix}

A minimal model for describing the physics of a HgTe quantum well was proposed by Bernevig, Hughes and Zhang\cite{Bernevig-Zhang-Science-2006}. It is derived using $\mathbf{k}\cdot\mathbf{p}$ theory, see e.g.~Ref.~\onlinecite{Bernevig-Zhang-Science-2006,Hankiewicz-NJP-2010,Qi-Zhang-review-RMP-2010} for further details. The BHZ model gives the following $4\times4$ Hamiltonian in 2D $(k_x,k_y)$-space
\begin{align}
H(k_x,k_y)=
\left(
\begin{array}{cc}
  h(\kk) & \mathbf{0} \\
  \mathbf{0} & h^\ast(-\kk)
\end{array}
\right),
\label{eq:BHZ-H-2D}
\end{align}
in the basis $\{|E+\rangle,|H+\rangle,|E-\rangle,|H-\rangle\}$. Here $|E\pm\rangle$ and $|H\pm\rangle$ are Kramer pairs of electron-like and hole-like states, respectively. The states $|E+\rangle$ and $|H+\rangle$ ($|E-\rangle$ and $|H-\rangle$) have positive (negative) angular momentum projection along the $z$ direction,  which is perpendicular to the 2D quantum well.  The $2\times2$ block in $H(k_x,k_y)$ is given by
\begin{align}
h(\kk)=
\left(
\begin{array}{cc}
  \e_k+M_k & A(k_x+ik_y)  \\
  A(k_x-ik_y) & \e_k-M_k 
\end{array}
\right),
\end{align}
where $\e_k=-D(k_x^2+k_y^2)$, $M_k=M_0-B(k_x^2+k_y^2)$ and $A$, $D$, $M_0$, $B$ are parameters. From this model, Bernevig \emph{et al.}\cite{Bernevig-Zhang-Science-2006} predicted that HESs exist for HgTe quantum wells thicker than a certain critical value of $63$\AA. Here we use the parameters for a quantum well thickness of $70$\AA~(i.e.~well within the TI regime)\cite{Qi-Zhang-review-RMP-2010}: 
\begin{align}
A &= 3.65\textrm{eV\AA},
\quad 
B = -68.6\textrm{eV\AA}^2,\nonumber\\ 
D &= -51.2\textrm{eV\AA}^2\
\textrm{and}\ 
M_0 = -0.01\textrm{eV}.
\label{eq:parameters-70AA-QW}
\end{align}

\begin{figure}
\includegraphics[width=0.4\textwidth,angle=0]{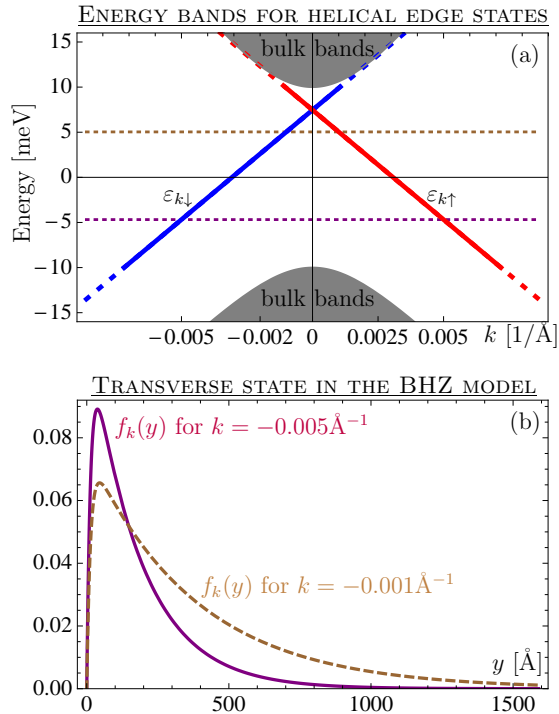}
\caption{(color online) (a). The energy dispersions $\e_{k\sigma}$ (red/blue full lines) for the HESs in the BHZ model for a 70\AA~thick HgTe quantum well Eq.(\ref{eq:parameters-70AA-QW}). The HES exist in the bulk band gap $E_g=2|M_0|$ and the bulk bands are shown in gray. (b). The transverse wave function Eq.(\ref{eq:trans-state}) in the BHZ model for $k=-0.001$\AA$^{-1}$(brown dashed line) and $k=-0.005$\AA$^{-1}$(purple full line), which have penetration depths of $\lambda_2^{-1}=378$\AA~and $\lambda_2^{-1}=177$\AA, respectively. The dotted (brown/purple) horizontal lines in (a) show the energy corresponding to these two transversal wave functions.}
\label{fig:trans-state-and-bands}
\end{figure}

A priori the Hamiltonian in Eq.(\ref{eq:BHZ-H-2D}) has periodic boundary conditions and therefore does not contain any edges. However, by introducing boundaries in the model Hamiltonian (\ref{eq:BHZ-H-2D}), it is possible to find edge states in the TI regime. One way to do this is by introducing hard wall boundary conditions and using the Peierls substitution $k_a=-i\p_a$ (for $a=x,y$)\cite{Zhou-edge-states-PRL-2008,Bihlmayer-Edge-states-in-Bi-films-PRB-2011}. The upper block $h(\kk)$ leads to one HES, while the lower block $h^\ast(-\kk)$ leads to its time-reversed Kramer partner. Therefore, the HESs are a mixture of the orbital states with either positive ($|E+\rangle$ and $|H+\rangle$) or negative ($|E-\rangle$ and $|H-\rangle$) total angular momentum projection. It is therefore that one can model the states as spin--$1/2$ in many situations, including the present one. In the limit of the boundaries being very well separated, the dispersions become exactly linear\cite{Bihlmayer-Edge-states-in-Bi-films-PRB-2011}, $\e_{k\op(\ned)}=\e_0+(-)\h v_0 k$, where the energy shift and velocity are given by $\e_0=-M_0D/B$ and $v_0=-\sqrt{B^2-D^2}|A|/(\h B)$, respectively, see Fig.~\ref{fig:trans-state-and-bands}(a). For a hard wall boundary at $y=0$ (and the TI at $y>0$), the (real and normalized) transverse wave function is found to be\cite{Zhou-edge-states-PRL-2008,Bihlmayer-Edge-states-in-Bi-films-PRB-2011}
\begin{subequations}
\label{eq:all-transverse-state-stuff}
\begin{align}
f_{k}(y)=
\sqrt{\frac{2\lambda_1\lambda_2(\lambda_1+\lambda_2)}{(\lambda_1-\lambda_2)^2}}
\left(e^{-\lambda_1 y}-e^{-\lambda_2 y}\right), 
\label{eq:trans-state}
\end{align}
where the $k$ dependence is in the parameters $\lambda_{1}$ and $\lambda_{2}$:
\begin{align}
\lambda_1
=&\frac{1}{\sqrt{B^2-D^2}} \left(\frac{|A|}{2}+\sqrt{W_{k}}\right),\\
\lambda_2
=&\frac{1}{\sqrt{B^2-D^2}} \left(\frac{|A|}{2}-\sqrt{W_{k}}\right).
\end{align}
Here we introduced
\begin{align}
W_{k}
=&\left[\frac{A^2}{4}-\frac{M_0}{B}(B^2-D^2)\right]
\nonumber\\
&+\frac{D|A|\sqrt{B^2-D^2}}{B}k
+(B^2-D^2)k^2. 
\end{align}  
\end{subequations}
Note that the transverse wave function vanishes at the boundary, $f_{k}(0)=0$, as required. It is evident from the form of the transverse state that the penetration depth or width is given by $W_y=\lambda_{2}^{-1}$ (since $\lambda_{2}<\lambda_{1}$). Figure \ref{fig:trans-state-and-bands}(b) shows this transverse state for the two different values of the wave vector $k$ corresponding to an energy difference as large as about half an energy band gap ($\sim10$meV).

In the main text, we use the transverse wave function with the parameters in Eq.(\ref{eq:parameters-70AA-QW}) as an example. We use it for numerical evaluation of the magnetization in Fig.~\ref{fig:pol-vs-eV} of the main text. Furthermore, in the averaging over the positions of the localized spins, we use that $\int_0^{\infty}dY_j[f_k(Y_j)]^4\simeq1/W_y$. This can be justified by using the BHZ transverse wave function Eq.(\ref{eq:trans-state}) such that 
\begin{align}
\int_0^{\infty}dY_j[f_k(Y_j)]^4
&=
\frac{3 \lambda_1 \lambda_2 (\lambda_1+\lambda_2)}{(3\lambda_1+\lambda_2)(\lambda_1+3\lambda_2)},
\end{align}
which in the limit $\lambda_{2}\ll\lambda_{1}$ gives the used result $\int_0^{\infty}dY_j[f_k(Y_j)]^4\simeq\lambda_2=1/W_y$. Moreover, the weak $k$ dependence of $W_y$ is neglected in the main text, which is seen to be reasonable both from the explicit form (\ref{eq:all-transverse-state-stuff}) and the examples given in Fig.~\ref{fig:trans-state-and-bands}(b).

%\bibliography{References-2012}

%Merlin.mbs v4.21 2009-07-09.
%

\end{document}